\documentclass[letterpaper,twocolumn,prl,aps,superscriptaddress,amsmath,amssymb,floatfix]{revtex4-1}
\usepackage{mathptmx}
\usepackage[latin9]{inputenc}
\usepackage{color}
\usepackage{mathrsfs}
\usepackage{amsmath}
\usepackage{amssymb}
\usepackage{graphicx}
\usepackage{esint}
\usepackage[unicode=true,
 bookmarks=true,bookmarksnumbered=false,bookmarksopen=false,
 breaklinks=false,pdfborder={0 0 1},backref=false,colorlinks=true]
 {hyperref}
\hypersetup{
 linkcolor=magenta,urlcolor=blue,citecolor=blue,pdfstartview={FitH},hyperfootnotes=false}

\makeatletter

\usepackage{textcomp}
\usepackage{epstopdf}

\usepackage{amsfonts}

\pdfpageheight\paperheight
\pdfpagewidth\paperwidth

\@ifundefined{textcolor}{}{%
 \definecolor{BLACK}{gray}{0}
 \definecolor{WHITE}{gray}{1}
 \definecolor{RED}{rgb}{1,0,0}
 \definecolor{GREEN}{rgb}{0,1,0}
 \definecolor{BLUE}{rgb}{0,0,1}
 \definecolor{CYAN}{cmyk}{1,0,0,0}
 \definecolor{MAGENTA}{cmyk}{0,1,0,0}
 \definecolor{YELLOW}{cmyk}{0,0,1,0}
}

\usepackage{xcolor}\usepackage{soul}
\definecolor{blue}{rgb}{0,0,1}
\definecolor{red}{rgb}{1,0,0}
\definecolor{green}{rgb}{0,1,0}

\makeatother

\begin{document}
\title{Quantum effects beyond mean-field treatment in quantum optics}
\author{Yue-Xun~Huang}
\affiliation{Key Laboratory of Quantum Information, Chinese Academy of Sciences,
University of Science and Technology of China, Hefei 230026, P. R. China.}
\affiliation{CAS Center For Excellence in Quantum Information and Quantum Physics,
University of Science and Technology of China, Hefei, Anhui 230026,
P. R. China.}
\author{Ming~Li}
\email{lmwin@ustc.edu.cn}

\affiliation{Key Laboratory of Quantum Information, Chinese Academy of Sciences,
University of Science and Technology of China, Hefei 230026, P. R. China.}
\affiliation{CAS Center For Excellence in Quantum Information and Quantum Physics,
University of Science and Technology of China, Hefei, Anhui 230026,
P. R. China.}
\author{Zi-Jie~Chen}
\affiliation{Key Laboratory of Quantum Information, Chinese Academy of Sciences,
University of Science and Technology of China, Hefei 230026, P. R. China.}
\affiliation{CAS Center For Excellence in Quantum Information and Quantum Physics,
University of Science and Technology of China, Hefei, Anhui 230026,
P. R. China.}
\author{Xu-Bo~Zou}
\affiliation{Key Laboratory of Quantum Information, Chinese Academy of Sciences,
University of Science and Technology of China, Hefei 230026, P. R. China.}
\affiliation{CAS Center For Excellence in Quantum Information and Quantum Physics,
University of Science and Technology of China, Hefei, Anhui 230026,
P. R. China.}
\author{Guang-Can~Guo}
\affiliation{Key Laboratory of Quantum Information, Chinese Academy of Sciences,
University of Science and Technology of China, Hefei 230026, P. R. China.}
\affiliation{CAS Center For Excellence in Quantum Information and Quantum Physics,
University of Science and Technology of China, Hefei, Anhui 230026,
P. R. China.}
\author{Chang-Ling~Zou}
\email{clzou321@ustc.edu.cn}

\affiliation{Key Laboratory of Quantum Information, Chinese Academy of Sciences,
University of Science and Technology of China, Hefei 230026, P. R. China.}
\affiliation{CAS Center For Excellence in Quantum Information and Quantum Physics,
University of Science and Technology of China, Hefei, Anhui 230026,
P. R. China.}
\date{\today}
\begin{abstract}
Mean-field treatment (MFT) is frequently applied to approximately
predict the dynamics of quantum optics systems, to simplify the system
Hamiltonian through neglecting certain modes that are driven strongly
or couple weakly with other modes. While in practical quantum systems,
the quantum correlations between different modes might lead to unanticipated
quantum effects and lead to significantly distinct system dynamics.
Here, we provide a general and systematic theoretical framework based
on the perturbation theory in company with the MFT to capture these
quantum effects. The form of nonlinear dissipation and parasitic Hamiltonian
are predicted, which scales inversely with the nonlinear coupling
rate. Furthermore, the indicator is also proposed as a measure of
the accuracy of mean-field treatment. Our theory is applied to the
example of quantum frequency conversion, in which mean-field treatment
is commonly applied, to test its limitation under strong pump and
large coupling strength. The analytical results show excellent agreement
with the numerical simulations. Our work clearly reveals the attendant
quantum effects under mean-field treatment and provides a more precise
theoretical framework to describe quantum optics systems.
\end{abstract}
\maketitle

\section{Introduction}

Mean-field treatment (MFT) is frequently applied to quantum many-body
systems to reduce the degrees of freedom for readily solvable evolution
equations. A typical application of MFT is demonstrated via the Ising
model and achieves great success in predicting phase transitions$\:$\citep{stanley1971mean}.
The main idea behind MFT is to focus on the average evolution of the
systems by discarding the quantum correlation and fluctuation~\citep{kadanoff2009more},
where the Bogoliubov inequality~\citep{bogolyubov2013method} serves
as the mathematical foundation. Within the realm of quantum optics,
MFT approximates the field of the optical modes to be complex numbers
by neglecting quantum statistics and has been widely adopted. Such
a treatment avoids the infinite dimension of number states and greatly
simplifies the system dynamics to semi-classical nonlinear equations.
It is demonstrated to be valid under weak coupling or strong drive.
In a more general form, the MFT is also applicable to a subsystem
by neglecting the quantum nature of part of the modes. The representative
examples are the manipulation of atom states and squeezed light generation
using strong coherent lasers, in which the strongly-pumped modes are
isolated from the rest of the system and solved independently following
the MFT.

Despite the great achievement in the application of MFT in quantum
optics, its validity is challenged for significant quantum correlations
and quantum fluctuations, which usually increase with the nonlinear
interaction strength. Another circumstance is the urgent need for
high-fidelity quantum information processors. The MFT on the control
field limits the fidelity of quantum gates since the control field
might entangle with the qubits. Even though these properties can be
expressed fully by tracking the density matrix in a large Hilbert
space, the computation complexity is beyond the capability of current
computers and it is not always necessary when the system does not
exhibit significant quantum features. Alternatively, it is efficient
to reveal these omitted quantum effects neglected by MFT and add these
elegant forms for a more accurate prediction of the system dynamics.
When the quantum correlations of a subsystem are neglected, decoherence
such as dephasing$\:$\citep{blais2004cavity} of the remaining system
could be expected by the reduced density matrix after tracing off
the subsystem. Even worse, such correlation can affect the expectation
values of the system and lead to the deviation from the prediction
by MFT. Several studies have focused on the side effects of the coupling
between a strong coherent field and a local quantum system, which
is common in many quantum optics experimental platforms. It is expected
that a high-intensity pump field can reduce the effect of dissipation~\citep{gea2002some,igeta2013fundamental,gea2010energy},
while it may actually lead to the strong interaction between the signal
and the pump modes~\citep{van2001classical}. To suppress or make
use of those correlation-induced phenomena beyond mean-field for high-fidelity
quantum device implement, we should have a clearer understanding of
how they depend on the experimental parameters thatoccurs with the
rapid development of quantum nonlinear optics devices. A case in point
is the advent of high-Q whispering gallery resonators~\citep{strekalov2016nonlinear},
which enables the strong coupling strength among different modes to
facilitate nontrivial quantum effects induced by correlation.

In this Letter, we offer the mean-field perturbation treatment (MFPT)
to extract the incidental quantum effects for an open quantum system
beyond the MFT. The general form of the incoherent dissipation and
parasitic coherent interaction are derived when a quantum system is
coupled to an MFT-applicable pump system. In addition, the indicators,
as the measure of the accuracy of MFT are put forward to reveal the
effect of quantum correlations on the system behavior. The theory
is introduced with a complete analysis of the quantum frequency conversion
process under a coherent pump, which agrees well with the simulation
and shows how the property of the coherent pump affects the fidelity
of the operation.

\section{Principle of Mean-Field Perturbation Treatment}

For an open quantum system, the Lindblad master equation offers a
general form to describe the system dynamics via \citep{lindblad1976generators}
$\dot{\rho}=\mathbb{L}\rho,$ where $\mathbb{L}$ is some generalized
Liouville super-operator and can be written as
\begin{equation}
\mathbb{L}=-i[H,.]+\sum_{\alpha}\kappa_{\alpha}\mathscr{L}_{d_{\alpha}}[.],\label{eq:FormL}
\end{equation}
where $H$ is the overall Hamiltonian for the system, $\mathscr{L}_{d_{\alpha}}[\rho]=2d_{\alpha}\rho d_{\alpha}^{\dagger}-d_{\alpha}^{\dagger}d_{\alpha}\rho-\rho d_{\alpha}^{\dagger}d_{\alpha}$
is the diagonalized Lindblad super-operator with the corresponding
decoherence rate $\kappa_{\alpha}$ and the subscript $\alpha$ denotes
different jump operator $d_{\alpha}$. In quantum optics systems,
the $H$ is composed of the bosonic field operators of different optical
modes, i.e. the annihilation operator $a$ and the creation operator
$a^{\dagger}$. Those operators are spanned over an infinite-dimension
Fock space, in which the equation is hardly solvable both analytically
and numerically. Therefore, MFT is applied by simply replacing some
of the bosonic operators with their expectation values in the simplified
mean-field Hamiltonian $H_{\mathrm{MFT}}$. In general, for an $N$-th
order term of the field operators in $H$ denoted as $\hat{N}=\widehat{N-M}\hat{M}$
with $M<N$, we can apply MFT by replacing the first term with its
expectation value $\langle\widehat{N-M}\rangle_{0}(t)$ and obtain
$\hat{N}_{\mathrm{MFT}}=\langle\widehat{N-M}\rangle_{0}(t)\hat{M},$
thus reducing the order from $N$ to $M$. Here, expectation values
$\langle\widehat{N-M}\rangle_{0}(t)$ are solved according to their
evolution under $H_{\mathrm{MFT}}$.

With the MFT, the system could be approximately solved with
\begin{equation}
\mathbb{L}_{\mathrm{MFT}}=-i[H_{\mathrm{MFT}},.]+\sum_{\alpha}\kappa_{\alpha}\mathscr{L}_{d_{\alpha}}[.].\label{eq:LMFT}
\end{equation}
As the generalized Liouville super-operators form an Abelian additive
group, we can still consider the full model by including the influence
of the dropped operators $\delta H=H-H_{\mathrm{MFT}}$ as
\begin{equation}
\delta\mathbb{L}=\mathbb{L}-\mathbb{L}_{\mathrm{MFT}}=-i[\delta H,.].\label{eq:deltaL}
\end{equation}
The essential idea of this work is to apply the perturbation of $\delta\mathbb{L}$
to the MFT, thus capturing the effects induced by the neglected quantum
correlations. According to the general perturbation theory~\citep{yi2000perturbative},
we obtain a set of equations to solve the evolution in power series
of $\delta\mathbb{L}$ as
\begin{equation}
\frac{d}{dt}\rho_{i}=\mathbb{\mathbb{L}}_{\mathrm{MFT}}\rho_{i}+\delta\mathbb{L}\rho_{i-1},\label{eq:MFPTTrunk}
\end{equation}
with $i\ge1$ denotes the order of the perturbation and $d\rho_{0}/dt=\mathbb{L}_{\mathrm{MFT}}\rho_{0}$
for the zeroth order. Eventually, the full solution of density matrix
is $\rho=\sum_{i=0}^{\infty}\rho_{i}$ with the normalized condition
$\mathrm{Tr}(\rho_{i})=0$ for $i\geq1$.

To provide a concrete discussion about the proposed mean-field perturbation
treatment (MFPT), we consider a general quantum system with a Hamiltonian
of $H=H_{\mathrm{I}}+H_{0}^{\mathrm{A}}+H_{0}^{\mathrm{B}}$, which
divided into two subsystems A and B, with the interaction between
the subsystems as $H_{\mathrm{I}}=\sum_{j=1}^{m}F_{j}^{\mathrm{A}}F_{j}^{\mathrm{B}}$,
which is decomposed into $m$ tensor products of operators in each
subsystem as $F_{j}^{\alpha}$ ($\alpha\in\{\mathrm{A},\mathrm{B}\}$).
We propose two approaches to apply the MFPT: (1) The\emph{ symmetric
}MFPT that applies the MFT to the interaction Hamiltonian symmetrically
to both subsystems as
\begin{equation}
H_{\mathrm{I},\mathrm{MFT}}=\sum_{j=1}^{m}\left(\langle F_{j}^{\mathrm{A}}\rangle_{0}F_{j}^{\mathrm{B}}+F_{j}^{\mathrm{A}}\langle F_{j}^{\mathrm{B}}\rangle_{0}\right)\label{eq:SymmetricMFPT}
\end{equation}
with $\delta H=\sum_{j=1}^{m}\delta F_{j}^{\mathrm{A}}\delta F_{j}^{\mathrm{B}}.$
(2) The\emph{ asymmetric }MFPT that only apply the MFT to one subsystem
$A$, thus
\begin{equation}
H_{\mathrm{I},\mathrm{MFT}}=\sum_{j=1}^{m}\langle F_{j}^{\mathrm{A}}\rangle_{0}F_{j}^{\mathrm{B}},\label{eq:AsymmetricMFPT}
\end{equation}
with $\delta H=\sum_{j=1}^{m}F_{j}^{B}\delta F_{j}^{A}$. It should
be remarked that for the symmetric MFPT, one can find the first-order
reduced density matrix for each subsystem are zero if the Lindblad
super-operators do not couple them together, i.e.
\begin{align}
\rho_{1}^{\mathrm{A}} & =\mathrm{Tr}_{\mathrm{B}}(\rho_{1})=0,\nonumber \\
\rho_{1}^{B} & =\mathrm{Tr}_{\mathrm{A}}(\rho_{1})=0.\label{eq:TrivialSolution}
\end{align}
 While for the asymmetric MFPT, only the reduced density matrix for
subsystem B is zero. A simple proof is provided in the Supplementary
Materials.

To reveal the quantum effects beyond the MFT, we can solve the perturbation
of the density matrix $\rho_{i}$ and Liouville super-operator $\delta\mathbb{L}$
by submitting Eq.$\:$(\ref{eq:MFPTTrunk}) into the master equation.
Here we consider a common case that only a subsystem ($A$) is strongly
driven by external coherent sources and weakly coupled with other
modes ($B$), thus we can applied the asymmetric MFPT. Beyond the
simple replacement of operators of subsystem $A$ by mean-field values,
the fluctuations of the modes in A could be treated as the noise of
the environment to $B$. Therefore, when we are focusing on the dynamics
of subsystem\textbf{ }B by tracing out $A$ from the whole density
matrix, the correlations between $A$ and $B$ might induce decoherences
and also shifting of the eigenspectrum of $B$. To understand such
effects, we first include all the interactions with the outer environments
explicitly in the total Hamiltonian in order to get rid of the Lindblad
super operators in $\mathbb{L}_{\textrm{MFT}}.$ Then we can turn
to the interaction picture with $\mathbb{L}_{\textrm{MFT}}$, after
which the equation for the reduced density matrix of subsystem B could
be deduced as
\begin{equation}
\frac{\partial\rho^{\mathrm{B}}}{\partial t}=\sum_{j=2}^{\infty}\frac{\partial\rho_{j}^{\mathrm{B}}}{\partial t},
\end{equation}
 where the evolution of $j$-th order perturbation component reads
\begin{align}
\frac{\partial\rho_{j}^{\mathrm{B}}}{\partial t_{j}} & =\left(-i\right)^{j}\int_{0}^{t_{j}}\ldots\int_{0}^{t_{2}}dt_{1}\ldots dt_{j-1}\nonumber \\
 & \times\mathrm{Tr}_{\mathrm{A}}\left[\delta H\left(t_{j}\right),\ldots,\left[\delta H\left(t_{1}\right),\rho_{0}\right]\right].\label{eq:DENMFPT}
\end{align}
For simplicity, we set the initial time to be zero, and the form of
time-dependent operators in the interaction picture can be deduced
via the Heisenberg-Langevin approach \citep{scully1999quantum}. To
the second-order perturbation, the evolution of subsystem $B$ follows
\begin{equation}
\frac{\partial\rho^{B}}{\partial t}\approx\frac{\partial\rho_{2}^{B}}{\partial t}=-\int_{0}^{t}dt_{1}Tr\left(\left[\delta H\left(t\right),\left[\delta H\left(t_{1}\right),\rho_{0}^{A}\otimes\right]\right]\right)\rho_{0}^{B}.
\end{equation}
To the limit of Markovian dynamics, since $\rho_{2}^{B}(t)$ is of
higher order, we replace $\rho_{0}^{B}$ with $\rho_{0}^{B}+\rho_{2}^{B}(t)$
and neglect the induced higher-order effects, we obtain the effective
extra Liouvillian due to the MFPT as
\begin{equation}
\frac{\partial\rho^{B}}{\partial t}\approx\mathbb{L}_{ex}\rho^{B}\label{eq:ExtraLiouvillian}
\end{equation}
 with $\mathbb{L}_{ex}=-\int_{0}^{t}dt_{1}Tr_{A}\left(\left[\delta H\left(t\right),\left[\delta H\left(t_{1}\right),\rho_{0}^{A}\otimes\right]\right]\right)$.
The super-operator $\mathbb{L}_{ex}$ can be classified into Hermitian
and non-Hermitian parts, corresponding to a modification of the Hamiltonian
of $B$ and the induced extra decoherence processes of $B$.

\section{Validity of MFT}

The MFPT provides a universal approach to include the incidental quantum
effects beyond the MFT by evaluating the influence of the fluctuations
in A. However, for the coupled two subsystems, there are also backaction
from B to A, and the modification of the evolution of B might also
contribute to A. Especially, when the coupling between the subsystems
is strong or the modes in subsystem B are excited to large amplitudes.
Considering an arbitrary operator $R$ and according to Eqs.~(\ref{eq:LMFT})-(\ref{eq:MFPTTrunk}),
we obtain
\begin{align}
\frac{d\langle R\rangle_{0}}{dt} & =-i\langle[R,H_{\mathrm{MFT}}]\rangle_{0}+\sum_{\alpha}\kappa_{\alpha}\langle\mathscr{L}_{d_{\alpha}}^{'}[R]\rangle_{0},\label{eq:EXPMFT}\\
\frac{d\langle R\rangle_{j}}{dt} & =-i\langle[R,H_{\mathrm{MFT}}]\rangle_{j}+\sum_{\alpha}\kappa_{\alpha}\langle\mathscr{L}_{d_{\alpha}}^{'}[R]\rangle_{j}\nonumber \\
 & \,\,-i\langle[R,\delta H]\rangle_{j-1}.\label{eq:EXPMFPT}
\end{align}
where $\mathscr{L}_{d_{\alpha}}^{'}[R]=2d_{\alpha}^{\dagger}Rd_{\alpha}-d_{\alpha}^{\dagger}d_{\alpha}R-Rd_{\alpha}^{\dagger}d_{\alpha}$
is the Lindblad operators in Heisenberg picture. We can deduce from
Eq.~(\ref{eq:EXPMFPT}) that if there are no cubic or higher-order
terms in the mean-field Hamiltonian and if $d_{\alpha}$ are all simple
field operators (first-order field operators), the dynamical equation
of any moment operator expectation $\langle R\rangle$ only couples
to other moment operators by a finite set of equations, which can
be solved analytically. Thus, by carefully selecting the mean-field
Hamiltonian, it is possible to extract the evolution of the expectation
values for the moment operators beyond MFT to any order readily via
Eq.~(\ref{eq:EXPMFPT}).

Therefore, we evaluate the validity of MFT by introducing an indicator
based on $\left\langle R\right\rangle _{j}$ with $j\geq1$ as a measure
of the accuracy of MFT. If the system admits a single physically steady
solution, we can solve the steady value of the moments by simply setting
the derivatives on the right side of the equations to zero~\citep{gomez2018perturbation}.
Following the idea of Ginzburg criterion~\citep{ginzburg1961some},
we are able to qualify the effectiveness of MFT by defining the $n$-th
order indicator as the absolute ratio of $n$-th order perturbation
to the sum of all lower order values, i.e.
\begin{equation}
\mathscr{I}_{n}[R]=\left|\frac{\langle R\rangle_{n,ss}}{\sum_{j=0}^{n-1}\langle R\rangle_{j,ss}}\right|.\label{eq:Indicator}
\end{equation}
If we expect MFT to work effectively, we shall make sure that any
order indicators of all operators to be uniformly small. While in
practice, we can simply check the first- or second-order indicators
($\mathscr{I}_{1}$ or $\mathscr{I}_{2}$) if the interaction Hamiltonian
is governed by an overall factor.

\section{Application of MFPT on quantum frequency conversion}

In this section, we verify the MFPT by analyzing a practical model,
i.e., quantum frequency conversion, which is usually realized by sum-frequency
generation (SFG) with Hamiltonian of
\begin{equation}
H=g(a^{\dagger}bc^{\dagger}+ab^{\dagger}c)+E_{a}(a^{\dagger}+a).\label{eq:full}
\end{equation}

\subsection{Asymmetric MFPT}

Treating modes $b$ and $c$ as the target quantum subsystem B and
mode $a$ as subsystem A, and applying the asymmetric MFPT, we have
\begin{equation}
H_{\mathrm{MFT}}=g(\alpha^{*}bc^{\dagger}+\alpha bc)+E_{a}(a^{\dagger}+a)
\end{equation}
with the perturbation $\delta H=g[(a^{\dagger}-\alpha^{*})bc^{\dagger}+(a-\alpha)b^{\dagger}c]$.
The three modes involved in the interaction experience energy decay
rates $\kappa_{a}$, $\kappa_{b}$ and $\kappa_{c}$. Since mode $a$
is dominant for $E_{a},\kappa_{a}\gg g$, thus the correlation between
mode $a$ and modes $b,c$ can be neglected, which is a necessary
condition of MFT. To simplify our analysis, we assume the dissipations
in system $B$ are small enough for an efficient conversion before
the loss of photons. We can easily find that the extra Liouvillian
{[}Eq.~(\ref{eq:ExtraLiouvillian}) {]} is relative to the integral
of $\langle\delta a(t)\delta a^{\dagger}(t_{1})\rangle_{0}(R(t)R^{\dagger}(t_{1})[.]-R^{\dagger}(t_{1})[.]R(t))$
and $\langle\delta a(t_{1})\delta a^{\dagger}(t)\rangle_{0}([.]R(t_{1})R^{\dagger}(t)-R^{\dagger}(t)[.]R(t_{1}))$
with $t\geq t_{1}$, and $R(t)=b^{\dagger}(t)c(t)$ can be solved
as
\begin{equation}
R(t)=\frac{1}{2}(X+We^{i2|G|t}+W^{\dagger}e^{-i2|G|t}),\label{eq:Rt}
\end{equation}
where $G=g\alpha$ is the stimulated linear coupling between modes
$b$ and $c$, $\bar{n}_{a}=|\alpha|^{2}=4E_{a}^{2}/\kappa_{a}$ is
the mean pump excitation number, $X=R-R^{\dagger}$, $Y=R+R^{\dagger}$,
$Z=[R,R^{\dagger}]=b^{\dagger}b-c^{\dagger}c$, and $W=\frac{1}{2}(Y-iZ)$
are the generalized Pauli operators for the subsystem B. Substituting
Eq.~(\ref{eq:Rt}) into the integral of $t_{1}$ and neglect the
high-frequency term under the assumption that $|G|\gg\gamma_{1}$,
we arrive at three extra Liouvillian in the interaction picture as
follows
\begin{align}
\mathbb{L}_{1} & =\gamma_{1}\mathscr{L}_{X}[.],\label{eq:SFGL1}\\
\mathbb{L}_{2} & =\gamma_{2}(\mathscr{L}_{Y}[.]+\mathscr{L}_{Z}[.])\label{eq:SFGL2}\\
\mathbb{L}_{3} & =-i[\Delta iX,.]\label{eq:SFGL3}
\end{align}
where $\gamma_{1}=\frac{g^{2}}{2\kappa_{a}}$, $\gamma_{2}=\frac{g^{2}\kappa_{a}}{4\left(\kappa_{a}^{2}+16g^{2}\bar{n}_{a}\right)}\approx\frac{1}{2}\gamma_{1}$,
and $\Delta=-\frac{2g^{2}}{\kappa_{a}^{2}+16g^{2}\bar{n}_{a}}|G|.$
We notice that Eq.~(\ref{eq:SFGL1}) and Eq.~(\ref{eq:SFGL2}) lead
to the dissipation of $X$, $Y$ and $Z$ with rates on the order
of $g^{2}$: $\mathscr{L}_{Z}$ leads to the degradation of diagonal
elements with a rate proportional to the photon number difference
of modes $b$ and $c$, while $\mathscr{L}_{Y}$ and $\mathscr{L}_{X}$
lead to the degradation of all elements. Since the rates of these
dissipations are all inverse to the dissipation rate $\kappa_{a}$
of the pump mode, it is beneficial to select a pump mode with large
dissipation for high-fidelity quantum frequency conversion. Surprisingly,
$\mathbb{L}_{1}$ is completely independent of the amplitude of the
pump laser and completely determined by the coupling rate and mode
dissipation. But for fixed linear coupling $G$, it will scale inversely
to the pump field, which is consistent with the previous results~\citep{gea2010energy}.
Since the rate in $\mathbb{L}_{2}$ is inverse to the averaged photon
number of mode $a$, $\mathbb{L}_{1}$ becomes dominant for strong
pump power. In addition to the decoherence effects, the eliminating
of mode $a$ also leads to the modification of the internal Hamiltonian
of system $B$, shown by Eq.~(\ref{eq:SFGL3}). The net effect of
$\mathbb{L}_{3}$ is to shift the Rabi oscillation frequency between
modes $b$ and $c$. Similar to $\mathbb{L}_{2},$it can be suppressed
with either a large dissipation rate or a large photon population.

\begin{figure}
\begin{centering}
\includegraphics[width=1\linewidth]{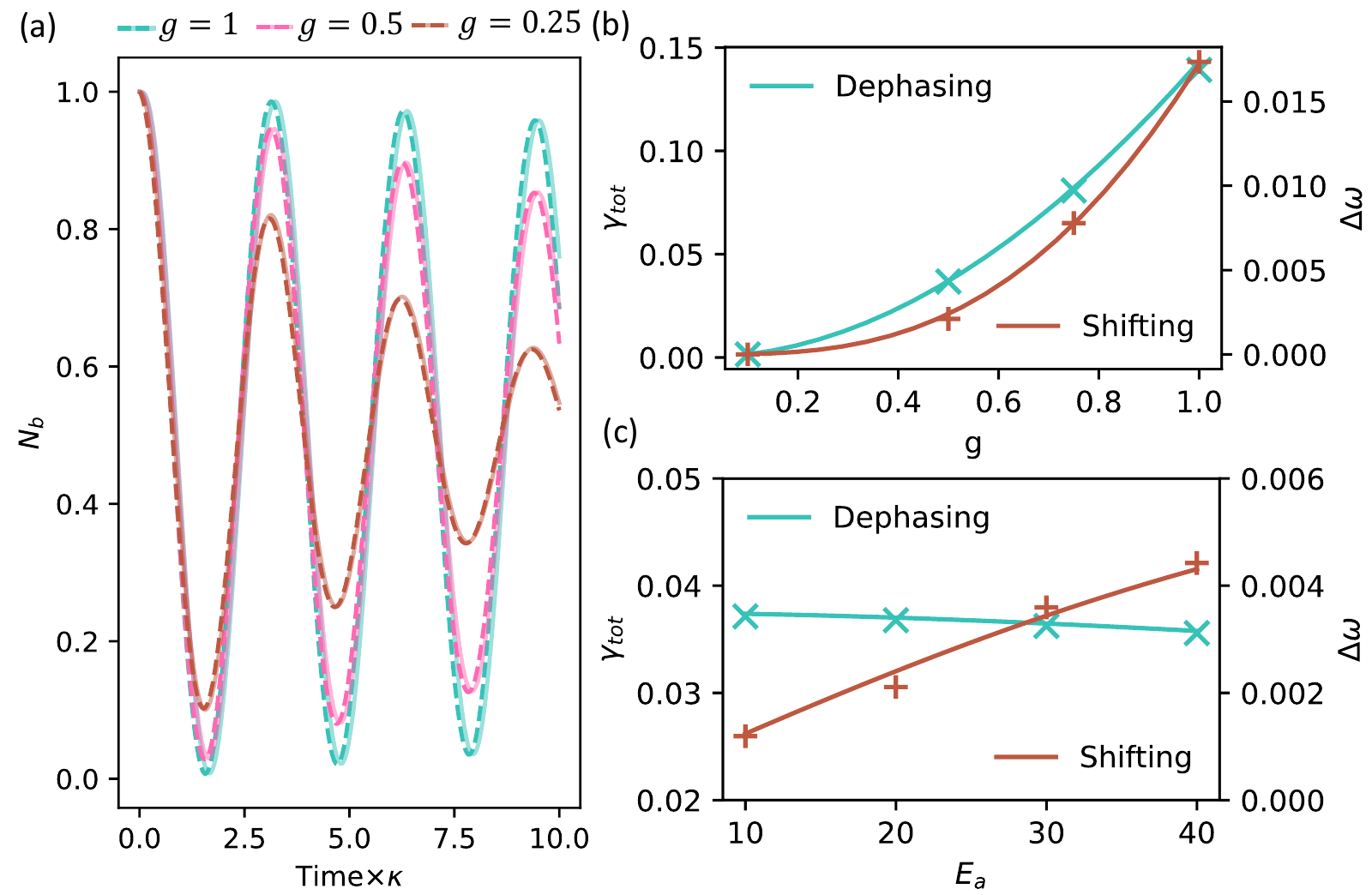}
\par\end{centering}
\caption{Comparison between numerical simulation results and analytical prediction
of quantum frequency conversion. Here, the parameters for all the
results are $\kappa_{a}=20$, $E_{a}=20$, and $g=0.5$. (a) The evolution
of photon population of mode $b$ under various $g$, with a fixed
$G$ via full model (solid lines) and $H_{\mathrm{MFT}}$ with extra
operators (dashed lines). (b) The dissipation rate and frequency shift
(solid line) under various $g$ and fixed $E_{a}$ and $\kappa_{a}$
compared with simulation results (cross). (c) The dephasing rate and
frequency shift (solid line) under various $E_{a}$ and fixed $g$
and $\kappa_{a}$ compared with simulation results (cross).}

\label{Fig1}
\end{figure}

The analytical result of MFPT is numerically verified by setting mode
$b$ with an initial single-photon state. In our MFPT framework, the
evolution of the photon population of mode $b$ can be $N_{b}=\frac{1}{2}(1+e^{-4(\gamma_{1}+\gamma_{2})t}\cos[(2G-2\Delta)t])$,
from which one see the Rabi frequency shift accompanied with a dissipation-induced
exponential decay {[}dashed lines in Fig.$\:$\ref{Fig1}{]}. Figures.$\:$\ref{Fig1}(b)-(c),
shows the relation between the dissipation rate, frequency shift,
and the coupling rate $g$ and driving strength $E_{a}$. All results
of our analytical results are in good agreement with numerical simulation
{[}solid lines in Fig.$\:$\ref{Fig1}{]} based on the full Hamiltonian
in Eq.$\:$(\ref{eq:full}). But the full model simulation costs a
much longer time to calculate. In our calculations, the averaged photon
number of mode $a$ is set large enough, thus only a single dissipation
rate of $\gamma_{1}$ is dominant.

\begin{figure}
\begin{centering}
\includegraphics[width=1\linewidth]{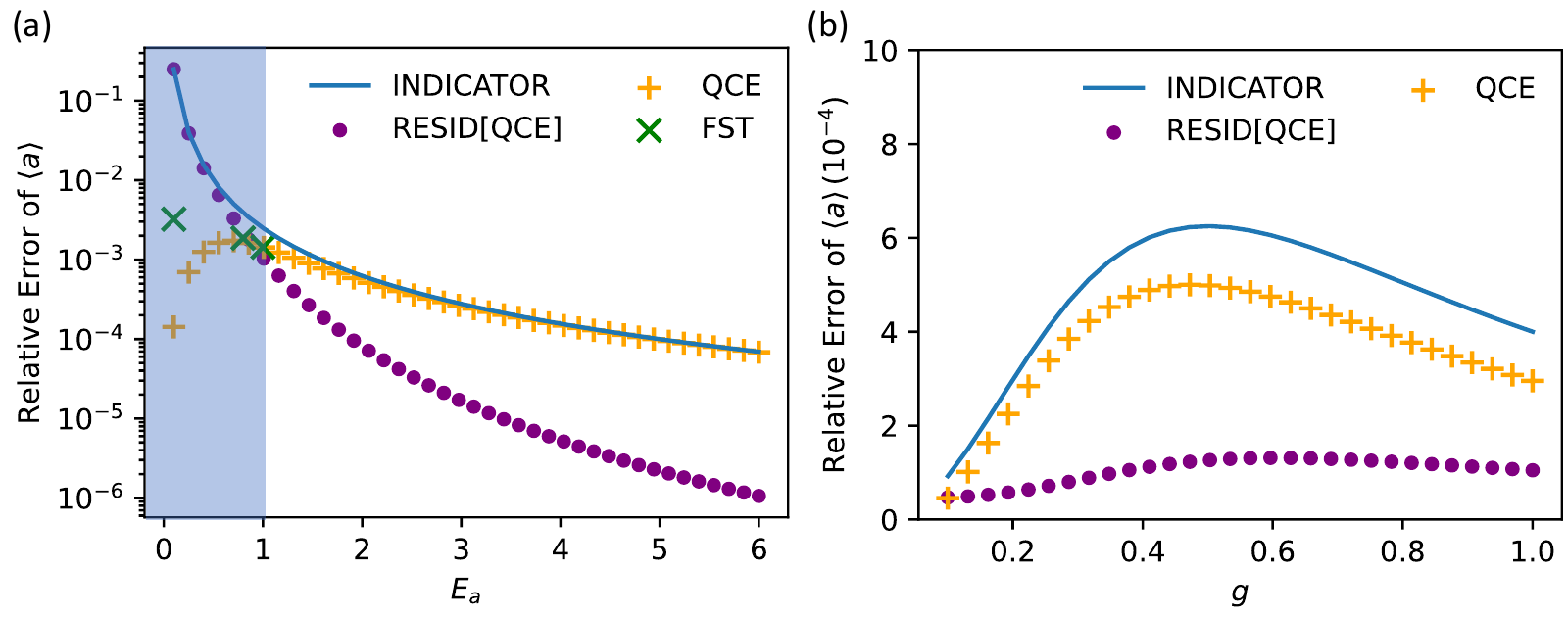}
\par\end{centering}
\caption{The indicator by the analytical prediction (blue solid line) and the
numerical simulation via quantum cluster expansion (QCE) method (orange
cross) and Fock state truncation (FST, green cross) under asymmetric
MFPT. The purple dot corresponds to the residual error between the
analytical results and the results from QCE. The relative error is
defined by $\textrm{Relative Error of }\langle R\rangle\triangleq\left|\frac{\langle R\rangle_{simu}-\langle R\rangle_{MFT}}{\langle R\rangle_{MFT}}\right|,$
where $\langle R\rangle_{simu}$ is the value given by full model
simulation and $\langle R\rangle_{MFT}$ is the value given by the
MFT. (a) The indicators under various drive strength of mode $a$
with fixed $G=-i$ and the shadow area indicates the parameter region
where asymmetric first-order MFPT may fail. (b) The indicators under
various $g$ with a fixed pump strength $E_{a}=2$. In both cases,
$\kappa_{a}=\kappa_{b}=\kappa_{c}=2$ and $E_{b}=0.1$ are fixed.}

\label{Fig2}
\end{figure}

The validity of the MFPT is also tested by considering the indicator
for the amplitude of subsystem A. As pointed out above, the mean-field
$\alpha$ of mode $a$ might not be solved independently if the coupling
strength $g$ or excitation numbers in mode $b,c$ is strong enough,
as the backaction from subsystem B would affect the mean-field of
mode $a$ in practical experiments. Considering the practical frequency
conversion experiments with a coherent signal input $E_{b}(b^{\dagger}+b)$
to mode $b$, and thus we have
\begin{align*}
\mathscr{I}[a]_{1} & =\frac{4E_{b}^{2}g^{2}}{\left(\frac{\kappa_{b}}{2}+\frac{2|G|^{2}}{\kappa_{c}}\right)^{2}\kappa_{a}\kappa_{c}}.
\end{align*}
As expected, larger $E_{a}$ ($G\propto gE_{a}$) and smaller $E_{b}$
lead to a smaller $\mathscr{I}[a]_{1}$, i.e. a better approximation
of $H_{\mathrm{MFT}}$. As shown in Fig.~\ref{Fig2}(a), we see that
the indicator decreases monotonously with $E_{a}$ if we fix $G$,
i.e. $g\propto1/E_{a}$. While for small pump power $E_{a}$ (shadow
area), the indicator grows extremely large and the MFT shall fail
here. The numerical results are obtained using the quantum cluster
expansion (QCE) method and Fock state truncation (FST) by Qutip~\citep{huang2021classical,johansson2012qutip}.
The deviation between the indicator and the results of QCE and FST
demonstrate such a statement. We can see from Fig.~\ref{Fig2}(b)
that the indicator fits well with the simulation with small discrepancy.
It justifies the argument that for strong coupling strength $g$,
a sufficiently large $E_{a}$ is necessary to ensure the accuracy
of traditional MFT.

\subsection{Symmetric MFPT}

\begin{figure}
\begin{centering}
\includegraphics[width=1\linewidth]{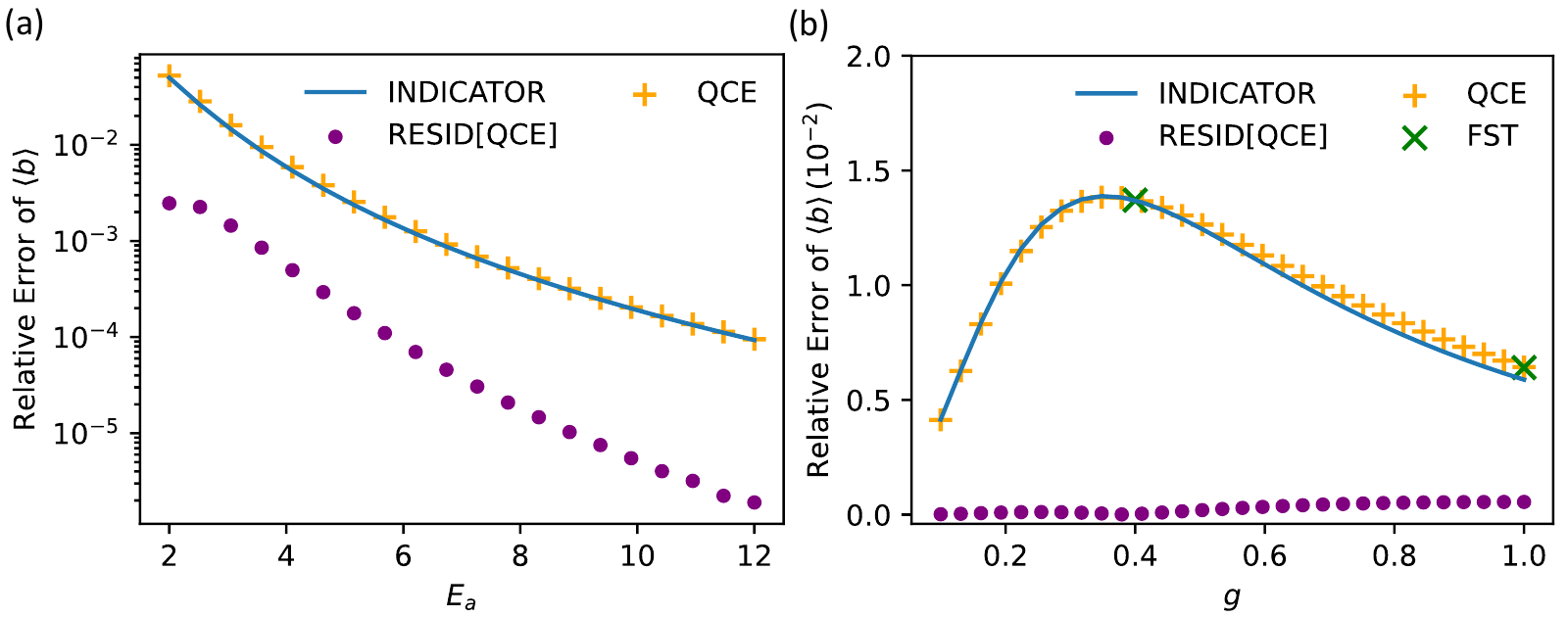}
\par\end{centering}
\caption{The indicator (solid line) compared to the simulation result of the
relative error via quantum cluster expansion (QCE) method~(orange
cross) and Fock state truncation (FST, green cross) under symmetric
MFPT. The purple dot corresponds to the residual between Indicator
and QCE. The relative error is defined by $\textrm{Relative Error of }\langle R\rangle\triangleq\left|\frac{\langle R\rangle_{simu}-\langle R\rangle_{MFT}}{\langle R\rangle_{MFT}}\right|,$
where $\langle R\rangle_{simu}$ is the value given by full model
simulation and $\langle R\rangle_{MFT}$ is the value given by the
MFT. (a) Results against $E_{a}$, with fixed coupling strength $g=1$.
(b) Results against coupling strength $g$, with fixed drive amplitude
$E_{a}=4$. In both cases, $\kappa_{a}=\kappa_{b}=\kappa_{c}=2$ and
$E_{b}=0.1$ are fixed.}

\label{Fig3}
\end{figure}

Lastly, we provide an example of symmetric MFPT. All the above analyses
are focusing on an asymmetric system with $E_{a}\gg E_{b}$ to give
a precise description of the system. In principle, adapting the asymmetric
MFPT can also describe the backward perturbation from subsystem B
to A when applying the perturbation theory to high orders. While the
calculation of the higher-order perturbations is cumbersome in practice
due to the existence of the first-order perturbation terms in subsystem
A. By treating the two subsystems on the same footing, symmetric MFPT
helps include all the mean-field influences between two subsystems
and extract the correction induced by the correlations to the higher-order,
which is quite suitable for investigating the case that $E_{b}$ is
comparable to $E_{a}$. For the symmetric MFPT
\begin{align}
H_{\mathrm{MFT}}= & g(\alpha^{*}bc^{\dagger}+\alpha bc)+E_{a}(a^{\dagger}+a)\nonumber \\
 & +g\left(a^{\dagger}\langle bc^{\dagger}\rangle_{0}+a\langle b^{\dagger}c\rangle_{0}\right)+E_{b}(b^{\dagger}+b),
\end{align}
the mean-field value of $a$ is the solution of the cubic equation
\begin{equation}
\alpha=-iE_{a}+\frac{E_{b}^{2}g^{2}\alpha}{(1+|\alpha|^{2}g^{2})^{2}},
\end{equation}
where we assume $\kappa_{a}=\kappa_{b}=\kappa_{c}=2$ for simplicity
in the following calculation. According to Eq.~(\ref{eq:TrivialSolution}),
the first order indicators for both subsystems are zero and we follow
the same process to calculate the second-order indicators as
\begin{align}
\mathscr{\mathscr{I}}[a]_{2} & =\left|\frac{-ig\langle bc^{\dagger}\rangle_{1,ss}}{\langle a\rangle_{0,ss}}\right|=0\\
\mathscr{I}[b]_{2} & =\mathscr{I}[c]_{2}=\left|\frac{2g^{2}}{(1+|\alpha|^{2}g^{2})(4-g^{2}|\alpha|^{2})}\right|.
\end{align}
It can be seen directly that the symmetric MFPT is a third-order approximation
for mode $a$ and thus the MFPT is efficient for subsystem A as long
as the pump is large enough. As shown in Fig.~\ref{Fig3}, the proposed
theory agrees quite well with the simulation of the relative error..

\section{Conclusion}

In summary, a general approach combining the MFT and the perturbation
theory is developed to extract the quantum correlation-induced effects
beyond the MFT in quantum optics systems. Upon the traditional MFT,
which separates the system into two subsystems and neglects the correlation
between them, we derive the additional decoherence effects and modification
of Hamiltonian to provide a more precise description of the quantum
system. Furthermore, indicators are provided to quantitatively measure
the accuracy of MFT. Via a practical quantum optics model of frequency
conversion, our approach is validated by comparing with numerical
simulations based on the full Hamiltonian. The indicators as well
as the extra super-operators derived by our approach help us design
the system parameters so that the unwanted effects could be suppressed
and optimal quantum operation fidelity could be achieved in experiments.
\textcolor{black}{Our approach provides a general framework of the
MFT, and the choice of MFT Hamiltonian} enjoys a high degree of freedom,
which may give rise to rich possibilities of extracting the appreciable
evolution beyond the mean-value. Further extension of the approach
to divide the system into more subsystems and the applications in
designing quantum devices are potential research topics for future
works.

\smallskip{}

\begin{acknowledgments}
Y.-X.H. thanks professor Wojdylo John Andrew in Nagoya University
for his inspiration for the general perturbation theory and Qianhui
Lu for her warm support all the time. This work was funded by the
National Key R\&D Program (Grant No.~2017YFA0304504), the National
Natural Science Foundation of China (Grants No.~11922411, No.~11904316,
and No.~12061131011). ML and CLZ was also supported by the Fundamental
Research Funds for the Central Universities (Grant No.~WK2470000031
and No.~WK2030000030), and the State Key Laboratory of Advanced Optical
Communication Systems and Networks. The numerical calculations in
this paper have been done on the supercomputing system in the Supercomputing
Center of University of Science and Technology of China.
\end{acknowledgments}


\begin{thebibliography}{10}
	\providecommand{\url}[1]{\texttt{#1}}
	\providecommand{\urlprefix}{URL }
	\providecommand{\eprint}[2][]{\url{#2}}
	
	\bibitem{stanley1971mean}
	H.~Stanley, Mean field theory of magnetic phase transitions, Introduction to
	Phase Transitions and Critical Phenomena  (1971).
	
	\bibitem{kadanoff2009more}
	L.~P. Kadanoff, More is the same; phase transitions and mean field theories,
	Journal of Statistical Physics \textbf{137}, 777 (2009).
	
	\bibitem{bogolyubov2013method}
	N.~N. Bogolyubov, \emph{A method for studying model Hamiltonians: a minimax
		principle for problems in statistical physics}, volume~43  (Elsevier 2013).
	
	\bibitem{blais2004cavity}
	A.~Blais, R.-S. Huang, A.~Wallraff, S.~M. Girvin, and R.~J. Schoelkopf, Cavity
	quantum electrodynamics for superconducting electrical circuits: An
	architecture for quantum computation, Physical Review A \textbf{69}, 062320
	(2004).
	
	\bibitem{gea2002some}
	J.~Gea-Banacloche, Some implications of the quantum nature of laser fields for
	quantum computations, Physical Review A \textbf{65}, 022308 (2002).
	
	\bibitem{igeta2013fundamental}
	K.~Igeta, N.~Imoto, and M.~Koashi, Fundamental limit to qubit control with
	coherent field, Physical Review A \textbf{87}, 022321 (2013).
	
	\bibitem{gea2010energy}
	J.~Gea-Banacloche, Energy constraints for quantum logic via nonlinear optical
	processes, Optics communications \textbf{283}, 719 (2010).
	
	\bibitem{van2001classical}
	S.~J. van Enk and H.~J. Kimble, On the classical character of control fields in
	quantum information processing, arXiv preprint quant-ph/0107088  (2001).
	
	\bibitem{strekalov2016nonlinear}
	D.~V. Strekalov, C.~Marquardt, A.~B. Matsko, H.~G. Schwefel, and G.~Leuchs,
	Nonlinear and quantum optics with whispering gallery resonators, Journal of
	Optics \textbf{18}, 123002 (2016).
	
	\bibitem{lindblad1976generators}
	G.~Lindblad, On the generators of quantum dynamical semigroups, Communications
	in Mathematical Physics \textbf{48}, 119 (1976).
	
	\bibitem{yi2000perturbative}
	X.~Yi, C.~Li, and J.~Su, Perturbative expansion for the master equation and its
	applications, Physical Review A \textbf{62}, 013819 (2000).
	
	\bibitem{scully1999quantum}
	M.~O. Scully and M.~S. Zubairy, \emph{Quantum optics}  (American Association of
	Physics Teachers 1999).
	
	\bibitem{gomez2018perturbation}
	E.~A. G{\'o}mez, J.~D. Casta{\~n}o-Yepes, and S.~P. Thirumuruganandham,
	Perturbation theory for open quantum systems at the steady state, Results in
	Physics \textbf{10}, 353 (2018).
	
	\bibitem{ginzburg1961some}
	V.~Ginzburg, Some remarks on phase transitions of the second kind and the
	microscopic theory of ferroelectric materials, Soviet Phys. Solid State
	\textbf{2}, 1824 (1961).
	
	\bibitem{huang2021classical}
	Y.-X. Huang, M.~Li, K.~Lin, Y.-L. Zhang, G.-C. Guo, and C.-L. Zou,
	Classical-to-quantum transition in multimode nonlinear systems with strong
	photon-photon coupling, arXiv preprint arXiv:2111.09557  (2021).
	
	\bibitem{johansson2012qutip}
	J.~R. Johansson, P.~D. Nation, and F.~Nori, Qutip: An open-source python
	framework for the dynamics of open quantum systems, Computer Physics
	Communications \textbf{183}, 1760 (2012).
	
\end{thebibliography}
\end{document}